\documentclass[amsmath,twocolumn,aps,prb]{revtex4}
\usepackage{bm}
\usepackage{color}
\usepackage{stackengine}
\usepackage{graphicx}
\newcommand{\bq}{\begin{equation}}
\newcommand{\ee}{\end{equation}}
\newcommand{\sgn}{\text{ sgn}}

\begin{document}

\title{Potential and spin-exchange interaction between Anderson impurities in graphene}

\author{M. Agarwal}
\affiliation{Department of Physics and Astronomy, University of Utah, Salt Lake City, UT 84112, USA}

\author{E. G. Mishchenko}
\affiliation{Department of Physics and Astronomy, University of Utah, Salt Lake City, UT 84112, USA}

\begin{abstract}
The effective interaction between resonant magnetic Anderson impurities in graphene, mediated by conduction electrons, is studied as a function of the strength of the onsite energy level of the impurities and the amplitude of coupling to conduction electrons. The sign and character of the interaction depend on whether the impurities reside on the same or opposite sublattices. For the same (opposite) sublattice, the potential interaction is attractive (repulsive) in the weak coupling limit with $1/R^3$ dependence on the distance; the interaction reverses sign and becomes  repulsive (attractive) in the strong coupling limit and displays $1/R$ behavior. The spin-exchange coupling is ferromagnetic (antiferromagnetic) at both large and small distances, but reverses sign and becomes anti-ferromagnetic (ferromagnetic) for intermediate distances. For opposite sublattices, the effective spin exchange coupling is resonantly enhanced at distances where the energy levels cross the Dirac points.
\end{abstract}

 \maketitle

\section{Introduction}

Doping novel two-dimensional materials with magnetic atoms is one of the active areas of research whose ultimate objective is to design systems with the desired magnetic properties. To better exploit an emerging magnetism in such doped materials, it is important to understand how magnetic impurities interact with each other.

Impurities in conventional three-dimensional metals induce famous charge (Friedel) and spin density (RKKY) oscillations of the conduction electron density, $\propto \cos(2k_{F}R)/(k_{F}R)^3$, in the long distance limit. In conventional two-dimensional electronic systems\cite{FK, BM} the amplitude of these oscillations decays inversely proportional to the square of the distance. One exception is graphene, a two dimensional material known for its remarkable electronic properties and a potential for applications\cite{CN}. Density oscillations in graphene in both intrinsic (undoped) and extrinsic (doped) limit decay as\cite{MM}  $\propto 1/R^3$, much like in three-dimensional systems.
The RKKY interaction between magnetic impurities in graphene has also been studied extensively\cite{Sar,BFS,BS,GKF,SS,Kog,PF,AFS}. The sign of the RKKY interaction for a bipartite lattice of intrinsic graphene at half-filling is dictated by the particle-hole symmetry and is anti-ferromagnetic (ferromagnetic) when the impurities reside on different (same) sublattices. This is found at all length scales\cite{Sar}. For example, RKKY exchange  coupling between spins of impurities located on the same sublattice has the following oscillatory behavior\cite{SS} $J_{AA}\propto - [1+ \cos(({\bf K}-{\bf K}')\cdot {\bf R})]/R^3$, where ${\bf K}$ and ${\bf K}'$ are the positions of two Dirac valleys in the reciprocal lattice. The coupling between spins on different sublattices, $J_{AB}$, has a similar oscillatory pattern, but the negative sign and the amplitude  that is three times larger than in the AA case.
\begin{figure}
        \centering
    	\includegraphics[scale = 0.45]{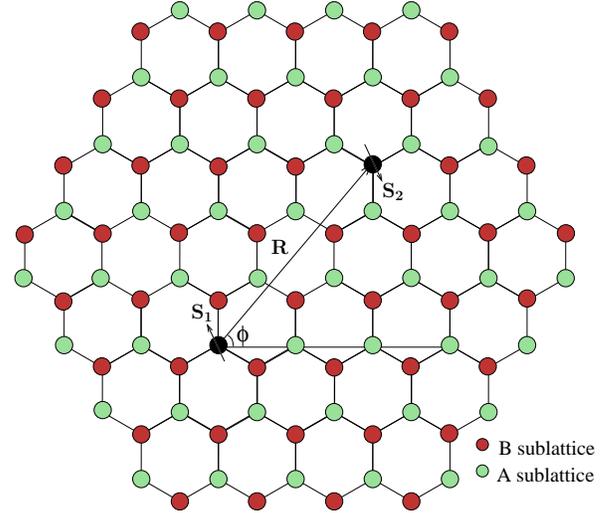}
	\caption{Graphene honeycomb structure consisting of two sublattices A(green) and B(red). Two impurities sitting on top of the carbon atoms with spin $S_{1}$ and $S_{2}$ are shown in black and are separated by vector $\bf {R}$. $\phi$ is the angle made by the $\bf {R}$ with zig-zag direction.}
	\label{fig 1}
\end{figure}

The above referenced studies considered interactions of  impurity atoms with the host material perturbatively. On the other hand, some adatoms (such as hydrogen) are better described by the Anderson model of a localized orbital hybridized with a conduction band of a host material. Such a model allows for a strong coupling of the localized orbital to conduction electrons.
In the present paper we consider two Anderson impurities with a low energy orbital $\epsilon_{o}$ hybridized with the $\pi$-band of graphene with some amplitude $\gamma$. We further assume that the orbital is below the Fermi energy (Dirac point) of undoped graphene, $\epsilon_o<0$, and that the Coulomb onsite energy $U_C$ is large enough, $\epsilon_o+U_C>0$, so that there is always an uncompensated spin $1/2$ associated with the impurity. Such assumptions work well for hydrogen adatoms, which have energy level close to the Dirac point of graphene\cite{WKL}. It is known that chemisorption of hydrogen atoms on graphene can indeed induce magnetic moments \cite{GGM}.

Magnetic applications of graphene would benefit from the ability to control magnetic moments. This, in turn, requires the knowledge of the magnitude and sign of the effective exchange coupling between dopants. Of particular interest is the behavior of resonant Anderson impurities, where the orbital $\epsilon_o$ resides close to the Dirac points\cite{MM,WKL}. This results in the  enhanced scattering of conduction electrons off the impurity\cite{MM}.

It is instructive to begin our analysis of the Anderson-type impurities with a discussion of potential impurities. Recent studies of impurity-impurity interaction in the case of substitution impurities  with an onsite potential $U$ have obtained an analytical expression exact in $U$\cite{SAL,LTM,AM}. In particular, in the large $U$ limit, interaction between impurities on the same sublattice is long range, $\propto 1/R$ (up to logarithmic factors), and is repulsive, in contrast to the weak $U$ limit where it decays as $1/R^3$ and is attractive. The interaction between impurities residing on opposite sublattices similarly reverses sign and changes behavior when the strength $U$ varies. Effectively, the Anderson impurity maps on the potential impurity model if one replaces the  onsite potential strength $U$ with the energy-dependent coupling parameter $\gamma(E)$, $U \to \gamma(E) = \frac{\gamma^2}{E - \epsilon_{o}}$. The weak potential impurity limit, analogous to the Anderson model with small $\gamma(E)$, maps onto the case of a large onsite energy $\epsilon_{o}$ and/or small amplitude $\gamma$ such that $\gamma(E) = -\gamma^2/\epsilon_{o}$ is an energy independent constant for most energies except $E\sim \epsilon_o$. As a result, the interaction of both types of impurities depends on the coordinate in a similar fashion,  $\propto U^2/R^3 \to \gamma^4/\epsilon_{o}^2 R^3$. With decreasing the distance $R$ between the impurities, the strong coupling limit is achieved when the on site energy $U$  becomes of the order of $v/R$. In this strong coupling limit,  the effective interaction energy is given by this very ratio $v/R$, since once $U$ drops out of the picture, there is only one remaining low-energy scale in the system. The sign of the interaction is repulsive (attractive) for impurities belonging to the same (different) sublattices. Below, in Section III we confirm that Anderson impurities in the resonant coupling regime demonstrate a similar $R$-dependence.

We then investigate the effective spin-spin exchange coupling $J_{\rm eff}(\bf R)$ between two Anderson impurities  in graphene and compare it with our recent results for substitutional potential impurities\cite{AM}. We limit our analysis to the lowest (second) order in the coupling $J$ between the localized spins and conduction electrons but explore a broad range of the  parameters $\epsilon_o$ and $\gamma$. The case of weak Anderson impurities yields,  $J_{\rm eff}({\bf R}) \propto J^2/R^3$, similar to effective spin coupling in the potential impurity case.
We then explore how $J_{\rm eff}({\bf R})$ behaves in the strong coupling limit. In our recent work\cite{AM}, we have shown that the effective spin exchange coupling between substitutional magnetic impurities can become resonantly enhanced at a specific distance where an impurity level crosses the Dirac point.  We find similar
enhancement for Anderson impurities for sufficiently large couplings $\gamma^2/\epsilon_o$.

This paper is organized as follows: In Section II, we discuss the energy levels of two Anderson impurities. In Section III and IV, we derive general expressions for the potential interaction and the spin exchange coupling between the impurities respectively and consider the limiting case of $\epsilon_o=0$.

\section{Energy levels of a two-impurity system}

We consider the tight binding model of $\pi$ electrons in graphene interacting with two Anderson impurities located at ${\bf r}_1 = 0$ and $\bf{r}_2 = {\bf R}$. In order to calculate the interaction between the impurities, we first determine the energy spectrum of the system.  The Hamiltonian of the system consists of the kinetic energy of electrons, the on site impurity energy level  $\epsilon_o$, and a coupling term describing hybridization of conduction band with the impurity states with amplitude $\gamma$,
\begin{align}
\label{Spinless Hamiltonian}
H_a = &t\sum_{{\bf r}}\sum_{i=1,2,3} \hat\psi^{\dagger}({\bf r})\hat\psi({\bf r+a}_i) + \epsilon_o \sum_{j=1,2}\hat d^{\dagger}({\bf r}_{j}) \hat d({\bf r}_{j}) \nonumber\\
& + \gamma \sum_{j=1,2}\Big( \hat d^{\dagger}({\bf r}_{j})\hat\psi({\bf r}_{j})+ \hat\psi^{\dagger}({\bf r}_{j}) \hat d({\bf r}_{j})\Big).
\end{align}
Here $t$ is the hopping integral,  $\hat\psi$ is electron operator of conduction electrons; $\hat \psi({\bf r})=\hat a ({\bf r})$ when ${\bf r}$ belongs to sublattice {\it A}, $\hat \psi({\bf r})=\hat b ({\bf r})$ when it belongs to sublattice {\it B}, and $\hat d$ is the operator of the localized impurity states. The index ${\it j}$ enumerates
the impurities. Using the Fourier representation for the electron operators, $\hat\psi ({\bf r})=\sqrt\frac{2}{N}\sum_{\bf k} \hat \psi({\bf k}) e^{i{\bf k}\cdot {\bf r}-iE t}$, we obtain from the equation of motion, $i\partial \hat \psi({\bf r},t)/\partial t =[\hat \psi({\bf r},t),H]$, the following system of coupled equations (for the AB impurity configuration),
\begin{eqnarray}
\label{a}
E \hat a({\bf k})&=&t({\bf k}) \hat b({\bf k})+\sqrt{\frac{2}{N}}\gamma d\hat(0),\\
\label{b}
E \hat b({\bf k})&=&t^*({\bf k}) \hat a({\bf k})+\sqrt{\frac{2}{N}}\gamma e^{-i{\bf k}\cdot{\bf R}}\hat d({\bf R}),\\
\label{c}
E\hat d(0) &=& \epsilon_o \hat d(0)+ \sqrt{\frac{2}{N}}\gamma \sum_{\bf k} \hat a({\bf k}),\\
\label{d}
E\hat d({\bf R})&=&\epsilon_o \hat d({\bf R})+ \sqrt{\frac{2}{N}}\gamma \sum_{\bf k} \hat b({\bf k}) e^{i{\bf k}\cdot{\bf R}},
\end{eqnarray}
where $t({\bf k})=t\sum_{i} e^{i{\bf k}\cdot{\bm a}_i}$ and $N$ is the total number of carbon atoms.
Eliminating $\hat d(0)$ and $\hat d({\bf R})$ gives,
\begin{eqnarray}
\label{e}
(E-\epsilon_o)[E\hat a({\bf k})-t({\bf k}) \hat b({\bf k})]&=&\frac{2\gamma^2}{N}\sum_{\bf k'} \hat a({\bf k'}),\nonumber\\
\label{f}
(E-\epsilon_o)[E\hat b({\bf k})-t^*({\bf k}) \hat a({\bf k})]&= &\frac{2\gamma^2}{N}\sum_{\bf k'} \hat b({\bf k'})e^{i({\bf k'}-{\bf k})\cdot{\bf R}}.\nonumber
\end{eqnarray}
Solving the above two equations yields the equation for the  localized impurity energy levels:
\begin{equation}
\label{AB}
\Bigl[1-\gamma^2\! \sum_{\bf k}A({\bf k},0)\Bigr]^2 = \gamma^4\! \sum_{\bf k}B({\bf k},{\bf R})\! \sum_{{\bf k}'}B(-{\bf k}',{\bf R}),
\end{equation}
where
\begin{equation*}
\left\{\begin{array}{l} A({\bf k},{\bf R}) \\ B({\bf k},{\bf R})\end{array} \right\} =
\frac{2e^{-i{\bf k}\cdot {\bf R}}}{N(E^2-|t({\bf k})|^2)(E-\epsilon_o)} \left\{\begin{array}{c} E \\ t({\bf k})\end{array} \right\}.
\end{equation*}
The poles in the above expression should be avoided in the usual way by replacing $E \to E+i\eta$.

Considering only low energy physics of the Hamiltonian, we expand momentum vector $\bf{k}$ near the two Dirac points, ${\bf k}={\bf K}_{\pm}+{\bf q}$. Summation over momentum vectors then gives,
\begin{equation}
\label{EnergyAB}
E = \frac{\epsilon_o \pm \alpha_0 v|\sin(\theta_{AB})|/R}{\alpha_0 (\ln|t/\epsilon_o| + i\frac{\pi}{2}) + 1}, \hspace{0.5cm} \alpha_0= \frac{\gamma^2 A_o}{\pi v^2}.
\end{equation}
Here $\theta_{AB}(\bm {R}) = \phi + \frac{2\pi R}{3 \sqrt3a} \cos\phi$, where $\phi$ is the angle measured from zig-zag direction as shown in Fig. \ref{fig 1}. The dimensionless constant $\alpha_0 \sim \gamma^2/t^2$ describes the strength of hybridization relative to the hopping integral. Importantly, one of the impurity levels in AB configuration can cross the Dirac point at a particular distance $R\sim \alpha_0v/\epsilon_o$. As we will see in Sec. IV, the spin exchange coupling between impurities residing on different sublattices can become resonantly enhanced at this distance where crossing occurs.

\section{Interaction energy: potential part}
The interaction energy of conduction electrons described by the Anderson Hamiltonian (\ref{Spinless Hamiltonian}) can be calculated using the following well-known quantum-mechanical identity,
\begin{multline}
\label{first_identity} \frac{\partial W}{\partial \gamma}= \left\langle \frac{\partial H}{\partial \gamma} \right\rangle
= \sum_{j=1,2}\left\langle \hat d^{\dagger}({\bf r}_{j})\hat\psi({\bf r}_{j})+ \hat\psi^{\dagger}({\bf r}_{j}) \hat d({\bf r}_{j})\right\rangle.
\end{multline}
This identity can be written in terms of electrons Green's function
\begin{equation}
\label{greensfunction} {\cal G}({\bf r},{\bf r'},t) =-i\langle T \hat\psi({\bf r},t)
 \hat\psi^\dagger({\bf r'},0)\rangle.
\end{equation}
Because according to Eqs.~(\ref{c}) and (\ref{d}), $\hat d({\bf r}_j) =\gamma \hat \psi({\bf r}_j)/(E-\epsilon_0)$, we obtain from Eq.~(\ref{first_identity}),
\begin{equation}
\label{secondidentity}
\frac{\partial W}{\partial \gamma} = - \frac{2i \gamma}{E-\epsilon_o}\sum_{j=1,2} {\cal G}({\bf r}_j,{\bf r}_j,t=0^-).
\end{equation}
The equation for Green's function in the  energy representation is
\begin{eqnarray}
\label{greens_equation} & E {\cal G}_{E}({\bf r},{\bf r'})- t\sum\limits_{i} {\cal G}_{E}({\bf r}+
{\bm a}_i,{\bf r'})-\gamma(E) \delta_{{\bf r},0} {\cal G}_{E}(0,{\bf r'})\nonumber\\
&-\gamma(E)\delta_{{\bf r},{\bf R}} {\cal G}_{E}({\bf R},{\bf r'})=\delta_{{\bf r},{\bf r'}},
\end{eqnarray}
where we introduced the shorthand,
\begin{equation}
\label{replacement}
\gamma(E)=\frac{\gamma^2}{E-\epsilon_o}.
\end{equation}
The solution of Eq.~(\ref{greens_equation}) has been found elsewhere\cite{LTM} for the  case of  two impurities with the onsite potential \textit{U}. Because the present case differs from that situation only by the  replacement  $U \to \gamma(E)$,  we can use the result of Ref.~\onlinecite{LTM},
\begin{equation}
\label{cal_G_o}
{\cal G}_{E}({\bf R},0) = G_E({\bf R},0)\frac{1+2T_EG_E(0,0)+T^{2}_EG_E^{2}(0,0)}{1-T^{2}_EG_E^{2}
(0,{\bf R})G_E({\bf R},0)},
\end{equation}
which expresses the two impurity Green's function ${\cal G}_{E}$ (for the electron propagation between two impurities) via the free electron Green's function $G_E$.

The  interaction energy (that part of $W$  that depends on the distance ${\bf R}$ between impurities) then follows from Eq. (\ref{secondidentity}):
\begin{equation}
\label{adatomW} W({\bf R})= 2i
\int\limits_{-\infty}^\infty\frac{dE}{2\pi}
\ln{\left(1-T^2_E G_E({\bf R},0) G_E(0,{\bf R}) \right)},
\end{equation}
where $T_E$ stands for the T-matrix,
\begin{equation}
\label{Tmatrixdefinition} T_E=\frac{\gamma(E)}{1-\gamma(E)G_E(0,0)}.
\end{equation}
The free electron Green's function evaluated at coinciding points, ${\bf r} = {\bf r}' = 0$ is
\begin{equation}
\label{Go}
G_E(0,0)=-\frac{E A_0}{\pi v^2}
\Bigl[\ln{\left(\frac{t}{|E|}\right)}+i\frac{\pi}{2} \! \sgn\,{E}\Bigr].
\end{equation}

Using now the fact that the time-ordered Green's functions do not have singularities in the first and third quadrants of the complex $E$-plane, and rotating the integration path counterclockwise by the angle $\pi/2$ so that it follows the imaginary axis, $E=i\omega$, we obtain the  expression for the interaction energy,
\begin{equation}
\label{adatomWimag} W({\bf R})= -2
\int\limits_{-\infty}^\infty\frac{d\omega}{2\pi}
\ln{\left(1-T^2_{i\omega} \Pi_{i\omega} ({\bf R}) \right)},
\end{equation}
where we introduced the following shorthand for the product of two Green's functions,
\begin{equation}
\Pi_{i\omega} ({\bf R}) = G_{i\omega}(0,{\bf R}) G_{i\omega}({\bf R},0).
\end{equation}
For the AA configuration of adatoms\cite{SAL,LTM},
\begin{equation}
\label{G_0R_G_R0_AA}
\Pi_{i\omega} ({\bf R})= -\frac{\omega^2A_o^2}{\pi^2v^4}K_0^2\left(\frac{|\omega|R}{v}\right)\cos^2\theta_{AA},
\end{equation}
where $K_{0}$ is the Macdonald function of the zeroth order and $\theta_{AA}(\bm {R}) = \frac{2\pi R}{3 \sqrt3a} \cos\phi$. For AB configuration the product is given by\cite{SAL,LTM},
\begin{equation}
\label{G_0R_G_R0_AB}
\Pi_{i\omega} ({\bf R})= \frac{\omega^2A_o^2}{\pi^2v^4}K_1^2\left(\frac{|\omega|R}{v}\right)\sin^2\theta_{AB},
\end{equation}
where $K_{1}$ is the Macdonald function of the first order and $\theta_{AB}$ is defined after Eq.~(\ref{EnergyAB}).

\begin{figure}
        \centering
    	\def\big{\includegraphics[scale = 0.36]{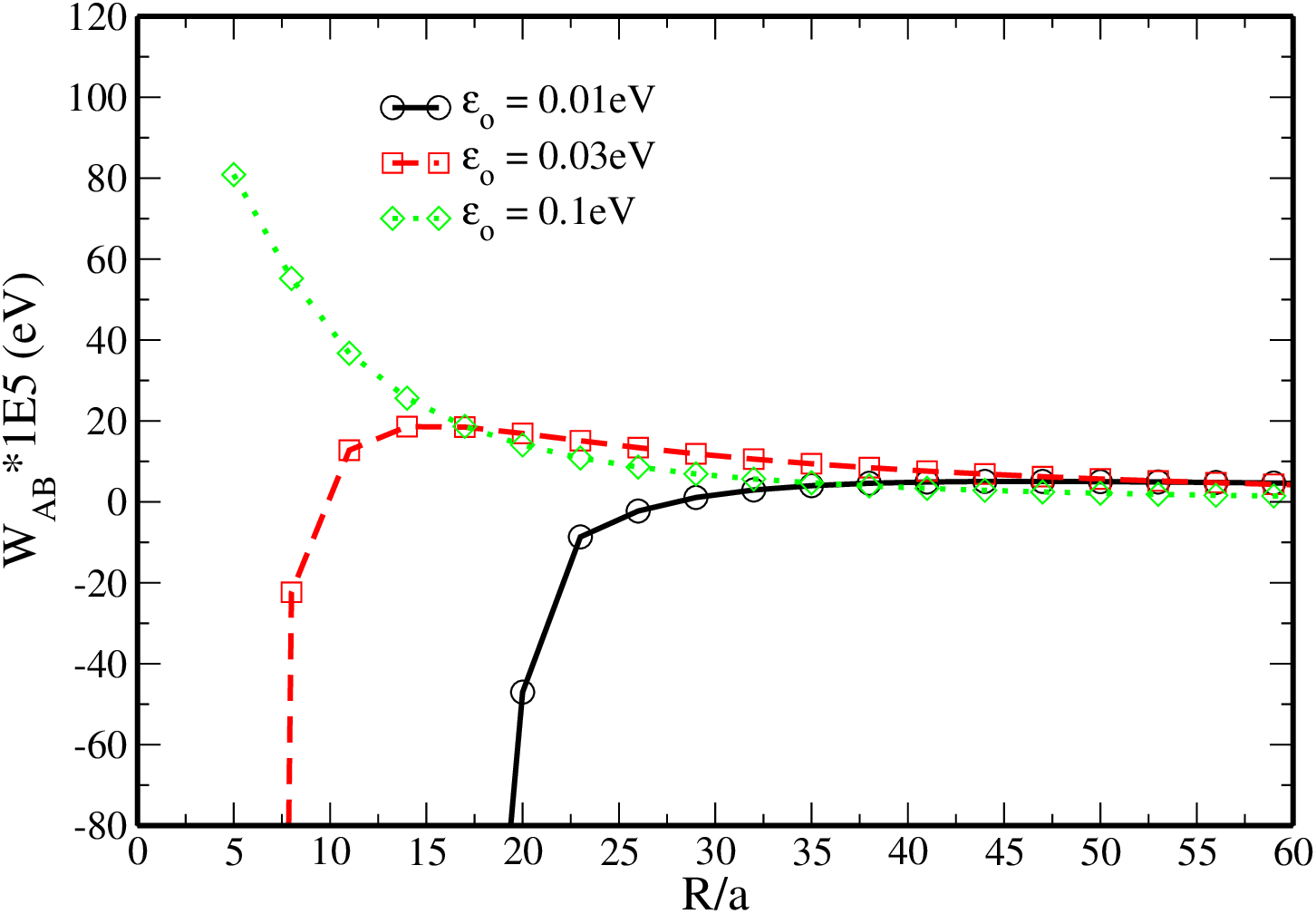}}
    	\def\little{\includegraphics[scale=0.13]{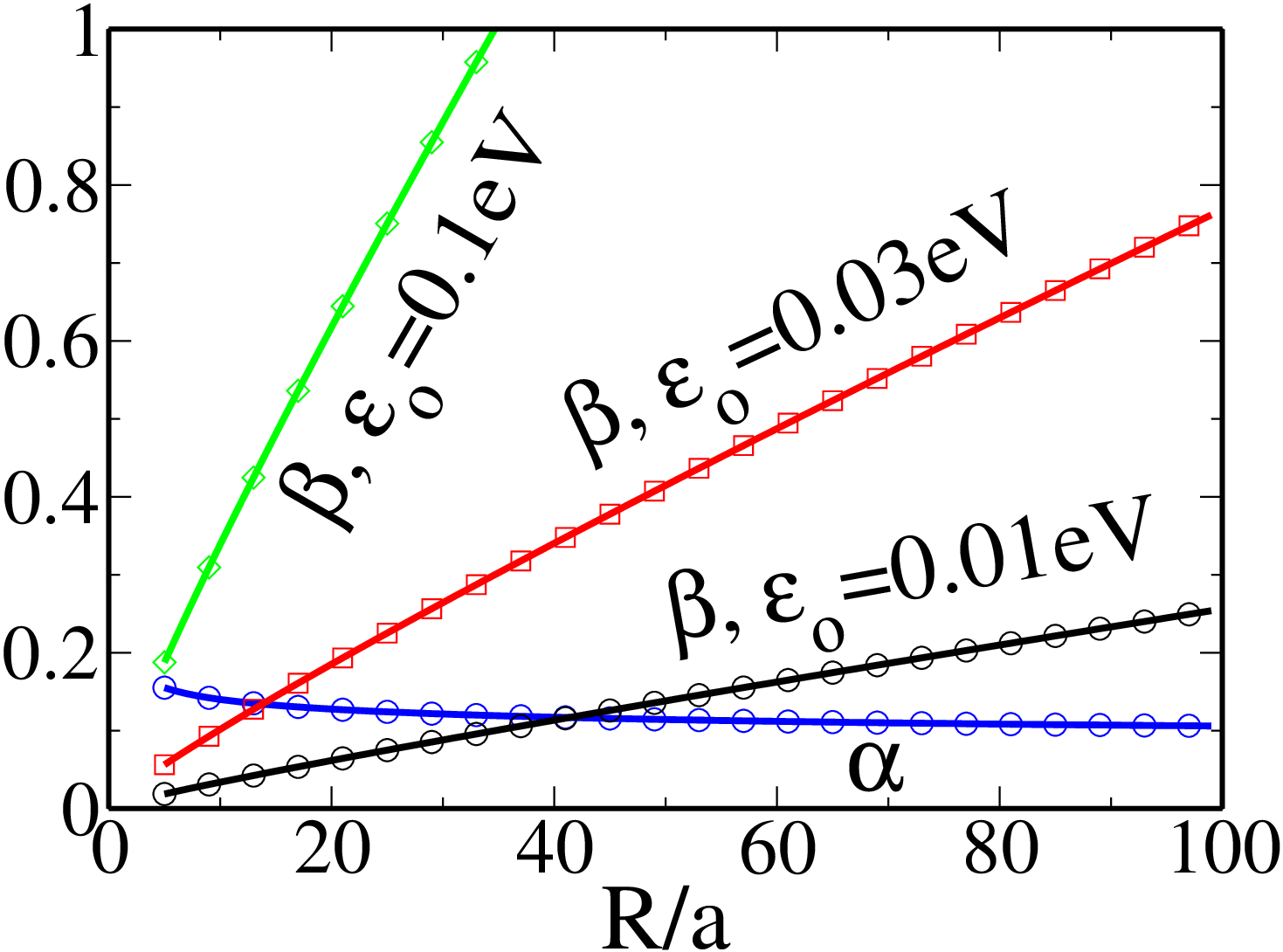}}
    	\def\stackalignment{r}
    	\topinset{\little}{\big}{14pt}{15pt}
	\caption{Interaction energy is plotted as a function of distance between the impurities \textit{R/a} in AB configuration for three different values of onsite energy $\epsilon_{o}$: 0.01, 0.03 and 0.1 eV. The coupling constant $\gamma = 1 $eV is same for all three plots in this figure. It is exact numerical plot of Eq. (\ref{W_AB}).}
	\label{fig 2}
\end{figure}
To make subsequent calculations of the energy $W(R)$ given by Eq. \ref{adatomWimag} more compact, let us introduce the two dimensionless parameters
\begin{equation}
\alpha = \frac{\alpha_0}{1+\alpha_0\ln(\frac{R}{a})}, \hspace{0.5cm}
\beta = \frac{R}{v} \frac{\epsilon_o}{1+\alpha_0\ln(\frac{R}{a})},
\end{equation}
namely, the renormalized impurity coupling strength $\alpha$ and the parameter $\beta$ that characterizes the location of the impurity level $\epsilon_o$ relative to the energy scale $v/R$ of the electron travel between the impurities.
With increasing the ``bare'' coupling $\alpha_0$, the renormalized $\alpha$ approaches a (distance-dependent) constant. Note that in the long-range limit $R\gg a$, to which the present theory only applies, $\alpha$ is always less than $1$. The parameter $\beta$, as we are about to see, describes the effective strength of the impurity with large $\beta > 1$ corresponding to the weak impurity limit  and $\beta <1$  to the strong coupling domain.

The potential interaction energy expression for impurities residing on different sublattices in terms of $\alpha$ and $\beta$ is given by,
\begin{equation}
\label{W_AB} W_{\it \! AB}({\bf R})= -\frac{2v}{R} \int\limits_{-\infty}^\infty\frac{dx}{2\pi} \ln{\left(1-\frac{\alpha^2 x^2 K^2_1\left(|x|\right) \sin^2\theta_{\it \! AB} }{(ix - \beta)^2} \right)}.
\end{equation}
In deriving the above expression we have used Eq. (\ref{G_0R_G_R0_AB}) for the product of two Green's function and the expression for the  $T$ matrix given by Eqs. (\ref{Tmatrixdefinition}) and (\ref{replacement}).
The integral in Eq.~(\ref{W_AB}) can be calculated in different limits of $\alpha$ and $\beta$.

(i) When the distance between the impurities is large enough so that $\beta \gg 1$, we can neglect $x$ in the denominator and calculate the remaining integral by expanding the logarithms over a small ratio $\alpha/\beta$ (as explained before, $\alpha<1$),
\begin{align}
\label{W_AB_shortrange}
 W_{\it \! AB}({\bf R}) & \approx \frac{-2v}{R} \int\limits_{-\infty}^\infty\frac{dx}{2\pi} \ln{\left(1-\left(\frac{\alpha }{\beta}\right)^2 \!x^2 K^2_1\left(|x|\right) \sin^2\theta_{\it \! AB}\right)} \nonumber\\
&\approx\frac{3\pi \alpha_0^2 } {16} \frac{v^3}{R^3 \epsilon_o^2} \sin^2\theta_{\it \! AB}.
\end{align}
This is simply the weak impurity limit, where interaction decays with the distance as $1/R^3$, just like in a case of a substitution impurity. The interaction is positive (repulsive) there. This regime is also realized when the impurity level $\epsilon_o$ is sufficiently far away from the Dirac point.

(ii) As the distance $R$ decreases or, alternatively, the energy level $\epsilon_o$ approaches the Dirac point, a situation of small $\beta \ll 1$
is eventually realized. (This condition means that the impurity level has the energy, $\epsilon_o \ll v/R$, i.e.  much smaller than the energy corresponding to the distance $R$). In the most interesting case of $\alpha \ll 1$, two scenarios can occur,  depending on how $\beta$ compares with $\alpha$. In the limit of $\alpha \ll \beta \ll 1$, one can still expand the logarithm in the integrand,
\begin{equation}\label{integralAB}
W_{\it \! AB}({\bf R}) \approx \frac{v \alpha^2 \sin^2\theta_{\it \! AB}}{\pi R} \int\limits_{-\infty}^\infty dx \frac{x^2 K^2_1\left(|x|\right)}{(ix - \beta)^2},
\end{equation}
even though it is no longer possible to neglect $ix$ in comparison to $\beta$ in the denominator.
Because this limit requires  small $\gamma$, the difference between two coupling constants becomes insignificant,  $\alpha \approx \alpha_0$, whereas $\beta \approx \epsilon_{o}R/v$. The remaining integral is calculated in Appendix \ref{Appendix A} to give,
\begin{equation}
\label{W_AB_longrange}
W_{\it \! AB}({\bf R})= \frac{\pi \alpha_0^2 v}{2R} \sin^2\theta_{\it \! AB} \left[1 + \frac{4\epsilon_o R}{\pi v}\ln\Big(0.89\frac{\epsilon_o R}{v}\Big)\right].
\end{equation}
The sign of the interaction remains the same as in Eq.~(\ref{W_AB_shortrange}) but the dependence on $R$ changes from $1/R^3$ to a long range $1/R$. As should be, the two expressions, Eq.~(\ref{W_AB_shortrange}) and Eq.~(\ref{W_AB_longrange}),  becomes of the same order when $\beta \sim 1$; this happens when $\epsilon_{o}\sim v/R$ .
\begin{figure}
        \centering
    	\def\big{\includegraphics[scale = 0.35]{{./images/wAB0.1eV_e0.05eV_compare}.eps}}
    	\def\little{\includegraphics[scale=0.14]{./images/insetWAB2.eps}}
    	\def\stackalignment{r}
    	\topinset{\little}{\big}{15pt}{15pt}
	\caption{Interaction energy is plotted as a function of distance between the impurities \textit{R/a} in AB           configuration for onsite energy $\epsilon_o = 0.05$ eV and coupling constant $\gamma = 0.1$ eV. The first plot labelled $exact$ is a result of the exact numerical integration Eq. (\ref{W_AB}). The other two plots labelled $1/R$ and $1/R^3$ are plotted using Eq. (\ref{W_AB_longrange}) and Eq. (\ref{W_AB_shortrange}) respectively.}
	\label{fig 3}
\end{figure}

In the second limit of $ \beta \ll \alpha \ll 1 $, where the impurity level energy $\epsilon_o$ is negligible in comparison with $v/R$, it is appropriate to ignore $\beta$ in the integrand in Eq.~(\ref{W_AB}),
\begin{align}
\label{firstW_AB_e0} W_{\it \! AB}({\bf R})& \approx -\frac{2v}{ R} \int\limits_{0}^\infty \frac{dx}{\pi}
\ln{[1+\alpha^2 \sin^2\theta_{\it \! AB} K^2_1(x)]}\nonumber\\ &\approx -\frac{2v}{ R} \int\limits_{0}^\infty \frac{dx}{\pi}
\ln{\left(1+\frac{\alpha^2 \sin^2\theta_{\it \! AB}}{x^2}\right)}.
\end{align}
Since the relevant values of $x$ in this integral are of the order of $\alpha$, neglecting $\beta$ in the original integral is justified.  Utilizing also that $\alpha$ is small, we
used the small-argument expansion of the Macdonald function, $K_{1}(x) \sim 1/x$. The remaining integral is straightforward to calculate using integration by parts and is equal to $\pi \alpha |\sin(\theta_{AB})|$. The expression of interaction energy is thus given by,
\begin{equation}
\label{secondW_AB_e0}
W_{\it \! AB}({\bf R})= -\frac{2 \alpha v}{R} |\sin\theta_{AB}|.
\end{equation}
Note that the presence of logarithmic term in $\alpha$ indicates the onset of multiple scattering of electrons off the impurities. This is the strong impurity limit where the interaction is attractive, in contrast to the weak impurity  limit. In the limit of $\alpha_0 \ln(R/a) \gg 1$,  we recover the expression found earlier in Refs.~\onlinecite{SAL,LTM} for strong potential impurities.

Fig. \ref{fig 2} illustrates the dependence of $W_{AB}$ on the distance between impurities for different values of onsite energy $\epsilon_{o}$: 0.1, 0.03 and 0.01 eV and coupling $\gamma$ = 1 eV. The inset plot in the figure is to explain the behavior of interaction energy with the help of impurity strength parameter $\beta$ and renormalized impurity coupling $\alpha$. It shows the variation of $\beta$ with distance $R/a$ for the above set of onsite energies $\epsilon_{o}$ along with $\alpha$ plotted for $\gamma = 1$ eV. For $\epsilon_{o} = 0.1 $ eV, $\beta$ remains greater than $\alpha$ for all values of $R/a$, hence the interaction energy is always repulsive. As we decrease $\epsilon_{o}$ to 0.03 eV and further down to 0.01 eV, we see the transition from weak coupling to strong coupling occurs leading to attractive interaction. It happens at the value of $R/a$ when $\beta \sim \alpha$. Fig. 3 on the other hand shows the dependence of $W_{AB}$ on the distance in different regime. Even though the potential interaction in this regime  is  repulsive, it changes from $1/R$ to $1/R^3$ dependence at $\beta \sim 1$.
\begin{figure}
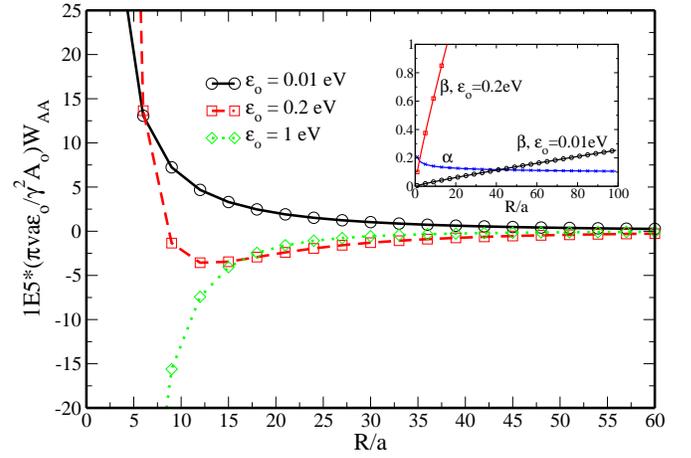

        \centering
    	\def\big{\includegraphics[scale = 0.35]{./images/W_AA.eps}}
    	\def\little{\includegraphics[scale=0.125]{./images/insetWAA1.eps}}
    	\def\stackalignment{r}
    	\topinset{\little}{\big}{14pt}{14pt}
	\caption{Interaction energy $W_{AA}$ is plotted as a function of distance between the impurities \textit{R/a} in AA configuration for three different values of onsite energy $\epsilon_{o}$: 0.01, 0.2 and 1 eV. The coupling constant $\gamma = 1 $ eV is same for all three plots in this figure. $W_{AA}$ is scaled by dimensionless ratio $ \pi v a \epsilon_{o}/\gamma^{2} A_{o}$. It is exact numerical plot of Eq. (\ref{W_AA}).}
	\label{fig 4}
\end{figure}
The  interaction energy for the two impurities residing on same sublattices is given by,
\begin{equation}
\label{W_AA} W_{\it \! AA}({\bf R})= \frac{-2v}{R} \int\limits_{-\infty}^\infty\frac{dx}{2\pi} \ln{\left(1+\frac{\alpha^2 x^2 K^2_0\left(|x|\right) \cos^2\theta_{\it \! AA} }{(ix - \beta)^2} \right)}.
\end{equation}
As in the  AB case, we calculate the above integral in different limits of $\alpha$ and $\beta$. In the weak impurity limit, $\beta \gg 1$, the integrand is simplified by neglecting $x$ in the denominator and the remaining integral can be calculated by expansion of log,
\begin{align}
\label{W_AA_shortrange} W_{\it \! AA}({\bf R}) &\approx \frac{-2v}{R} \int\limits_{-\infty}^\infty\frac{dx}{2\pi} \ln{\left(1+\left(\frac{\alpha }{\beta}\right)^2 x^2 K^2_0\left(|x|\right) \right)} \nonumber\\
&\approx -\frac{\pi  \alpha_o^2 } {16 }\frac{v^3}{R^3 \epsilon_o^2}\cos^2\theta_{\it \! AA}.
\end{align}
Because the integral converges over $x \sim 1 \ll \beta$, neglecting $x$ in comparison to $\beta$ in the denominator is justified. The interaction in the AA configuration in weak impurity limit is attractive, in contrast to the repulsive interaction  in the AB case and is three times smaller in magnitude.

When  $\beta \ll 1$, similarly to the AB case,  two limits arise.
 (i) For $\alpha \ll \beta \ll 1$, we are still justified to expand the logarithm (but not to neglect $x$ in the denominator),
\begin{equation}\label{integralAA}
W_{\it \! AA}({\bf R}) \approx \frac{v \alpha^2 \cos^2\theta_{\it \! AA}}{\pi R} \int\limits_{-\infty}^\infty dx \frac{x^2 K^2_0\left(|x|\right)}{(x + i\beta)^2}.
\end{equation}
The above integral is calculated in Appendix \ref{Appendix A} and the expression of interaction energy is given by,
\begin{align}
\label{W_AA_longrange}
W_{\it \! AA}({\bf R})= &\frac{v \alpha^2 \cos^2\theta_{\it \! AA}}{\pi R} \Big(\frac{\pi}{2} - \frac{\pi^2 }{2}\frac{\epsilon_o R}{v} -     \nonumber\\
&\frac{2\epsilon_o R}{v}\ln^2\Big(\frac{\epsilon_o R}{v}\Big) -  \frac{2\epsilon_o R}{v}\ln\Big(\frac{\epsilon_o R}{v}\Big)\Big).
\end{align}

\begin{figure}
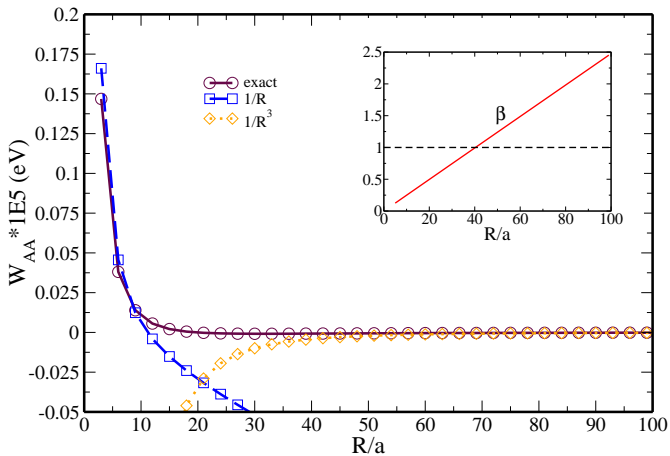

        \centering
    	\def\big{\includegraphics[scale = 0.35]{{./images/wAA0.1eV_e0.05eV_compare}.eps}}
    	\def\little{\includegraphics[scale=0.14]{./images/insetWAB2.eps}}
    	\def\stackalignment{r}
    	\topinset{\little}{\big}{15pt}{17pt}
	\caption{Interaction energy is plotted as a function of distance between the impurities \textit{R/a} in AA configuration for onsite energy $\epsilon_o = 0.05$ eV and coupling constant $\gamma = 0.1$ eV. The first plot labelled $exact$ is a result of the exact numerical integration Eq. (\ref{W_AA}). The other two plots labelled $1/R$ and $1/R^3$ are plotted using Eq. (\ref{W_AA_longrange}) and Eq. (\ref{W_AA_shortrange}) respectively.}
	\label{fig 5}
\end{figure}

(ii) In the remaining limit of $\beta \ll \alpha \ll 1$, we can neglect $\beta$ in denominator of the integral in Eq.~(\ref{integralAA}) and obtain,
\begin{equation}
\label{firstW_AA_e0} W_{\it \! AA}({\bf R}) =  \frac{\pi \alpha^2 v }{2R} \cos^2\theta_{\it \! AA}.
\end{equation}
We see that in the strong impurity limit the interaction is repulsive, in contrast to the weak impurity  limit Eq.~(\ref{W_AA_shortrange}) where it is attractive.

Fig. \ref{fig 4} illustrates the dependence of $W_{AA}$ on the distance between impurities for different values of the onsite energy $\epsilon_{o}$: 1, 0.2, 0.01 eV and coupling $\gamma = 1$ eV. The inset plot in the figure is to explain the behavior of interaction energy with the help of impurity strength parameter $\beta$ and renormalized impurity coupling $\alpha$. It shows the variation of $\beta$ with distance $R/a$ for the above set of onsite energies $\epsilon_{o}$ along with $\alpha$ plotted for $\gamma = 1$ eV. For $\epsilon_{o} = 1 $ eV, $\beta$ remains greater than one for all values of $R/a$, hence the interaction energy is always attractive. As we decrease $\epsilon_{o}$ to 0.2 eV and further down to 0.01 eV, we see the transition from weak coupling to strong coupling occurs leading to repulsive interaction. It happens at the value of $R/a$ when $\beta \sim \alpha$. Fig. 5, on the other hand shows the dependence of $W_{AA}$ on the distance in different regime. The potential interaction in this regime changes from attractive, $1/R^3$, to repulsive, $1/R$, at $\beta \sim 1$.

\section{Interaction energy: spin-dependent part}

To describe a spin-dependent part of the interaction between Anderson magnetic impurities in graphene, we add to our Hamiltonian the spin term,
\begin{equation}
H = H_a + H_{sp},
\end{equation}
where $H_a$ is given by Eq. (\ref{Spinless Hamiltonian}) and
\begin{equation}
\label{spin part}
H_{sp} = J{\bf S_1}\cdot \hat\psi_{\alpha}^{\dagger}(0)\hat{\bm \sigma}_{\alpha\beta}\hat\psi_{\beta}(0) + J {\bf S_2\cdot\hat\psi_{\alpha}^{\dagger}({\bf R})\hat{\bm \sigma}_{\alpha\beta}}\hat\psi_{\beta}({\bf R}).
\end{equation}
contains two short-range exchange interactions between spins of impurities,  $\textbf{S}_{1}$,  $\textbf{S}_{2}$, and those of conduction electrons described by the Pauli matrices  $\hat{\bm \sigma}$. The exchange coupling $J$ is assumed to be small compared with both $t$ and $\gamma$. As a result of the coupling to conduction electrons, there appears the effective coupling of impurity spins,
\begin{equation}
\label{effective coupling}
H_{\rm eff} =  J_{\rm eff}({\bf R})\, {\bf S}_1 \cdot {\bf S}_2.
\end{equation}
The effective exchange constant $J_{\rm eff}$ can be obtained from the already familiar method of differentiation with respect to the coupling parameter $J$,
\begin{multline}
\label{second_identity} \frac{\partial J_{\rm eff}}{\partial J}
{\bf S}_1 \cdot {\bf S}_2 = \left\langle \frac{\partial
H}{\partial J}\right\rangle= \sum_{j=1,2} {\bf S}_j \cdot\left\langle
\hat\psi^{\dagger}({\bf r}_j)\hat {\bm \sigma} \hat \psi({\bf r}_j)
\right\rangle.
\end{multline}
The expectation values in Eq.  (\ref{second_identity}) should be calculated to the lowest (first) order in the Hamiltonian (\ref{spin part}).  This yields,
\begin{equation}
\label{partial_Jeff}
 J_{\rm eff}= -2iJ^2\int\limits_{-\infty}^{\infty} \frac{dE}{2\pi \hbar}
                                   {\cal G}_{E}({\bf R},0){\cal G}_{E}(0,{\bf R}).
\end{equation}
The last expression can be simplified further by expressing Green's functions via free electron Green's functions, Eq.~(\ref{cal_G_o}). At last,  rotating the integration path counterclockwise by the angle $\pi/2$, $E = i\omega$, we obtain,
\begin{equation}
\label{Jeff} J_{\rm eff}= \frac{J^2}{\pi \hbar}
\int\limits_{-\infty}^{\infty} d\omega
\frac{\Pi_{i\omega} ({\bf R})}{[(1- \gamma ({i\omega})G_{i\omega}(0,0))^2- \gamma^2({i\omega})\Pi_{i\omega} ({\bf R})]^2},
\end{equation}
Having obtained the general expression for the effective spin exchange coupling for two impurities in AA or AB configuration, we can now proceed with evaluating it for different limits of the coupling strength $\gamma$ and  the position of the impurity level $\epsilon_{o}$.

\subsection{AB configuration}
\begin{figure}
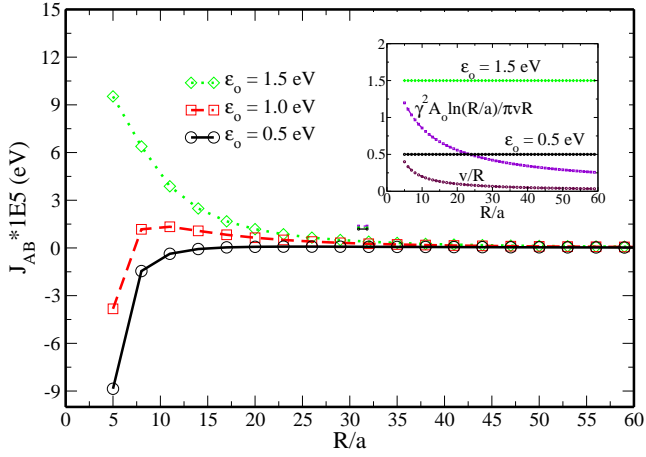

        \centering
    	\def\big{\includegraphics[scale = 0.35]{./images/wout_reson3eV.eps}}
    	\def\little{\includegraphics[scale=0.13]{./images/insetJAB1.eps}}
    	\def\stackalignment{r}
    	\topinset{\little}{\big}{14pt}{15pt}
	\caption{Effective spin coupling is plotted as a function of distance between the impurities $R/a$ in AB configuration for three different values of onsite energy $\epsilon_{o}$: 0.5, 1.0 and 1.5 eV. The coupling constant $\gamma = 3 $eV is same for all three plots in this figure. It is exact numerical plot of Eq. (\ref{J_AB}).}
	\label{fig 6}
\end{figure}
We begin by calculating spin-exchange coupling in AB configuration using Eq. (\ref{G_0R_G_R0_AB}) for the product of two Green's functions $\Pi(\bf{R})$ and $\gamma(i \omega) = \gamma^2/(i\omega - \epsilon_{o})$. Green's function at coinciding points $G_{i\omega}(0,0)$ is given by Eq. (\ref{Go}). To simplify calculations further let us introduce dimensionless distance $\rho = \epsilon_{o}R/v\alpha_{0}$ as (here, $\alpha_{o} = \gamma^2A_{o}/\pi v^2$, defined earlier in Eq. (\ref{EnergyAB})),
\begin{align}
\label{J_AB}
&J_{\rm eff}^{AB}({\bf R})= \frac{J^2}{\pi^3 \hbar \alpha_{0}^4}\frac{A_o^2 }{v R^3 } \sin^2\!\theta_{AB} \times\nonumber\\
&\int\limits_{-\infty}^{\infty}  \frac{dx x^2K_1^2(|x|)(ix-\alpha_0\rho)^4 }
{\left[ \left(\rho -ix\Big(\frac{1}{\alpha_{0}}+\ln{\left[\frac{R}{a|x|}\right]\Big)}
\right)^2  -\sin^2\!\theta_{AB}x^2K_1^2(|x|)\right]^2}.
\end{align}
Below we consider the dependence of the effective exchange constant on the distance $R$  (represented by the dimensionless variable $\rho$), as predicted by Eq.~(\ref{J_AB}).

At both large and small distances $R$, we obtain the same power-law dependence,  $ J_{AB}\propto 1/R^3$. Indeed, for very large $\rho$, the integral in Eq.~(\ref{J_AB}) is equal to $\alpha_0^4 \int_{-\infty}^{\infty} dx x^2K_1^2(|x|) =3\pi^2\alpha_0^4/16$, so that the exchange coupling becomes,
\begin{equation}
\label{J_AB_e0}
 J_{\rm eff}^{\! AB}({\bf R})= \frac{3J^2}{16\pi \hbar} \frac{ \! A_o^2}{v R^3}\sin^2\!{\theta_{\! AB}}.
\end{equation}
The applicability of this expression follows from the observation that $x\sim 1$ contribute to the integral and that $\rho$ must be large enough compared with $\frac{1}{\alpha_{0}}+\ln \frac{R}{a}$, or in terms of the actual distance, $R\gg (v/\epsilon_{o} )(1+\alpha_0 \ln \frac{R}{a})$.

On the other hand, at small distances one can simply set  $\rho\to 0$ in the integral. This again reveals the $R^{-3}$ dependence as the integral still converges at $x\sim 1$. Provided that
$\frac{1}{\alpha_{0}}+\ln \frac{R}{a} \gg 1$, one can also neglect the  $K_1^2(|x|)\sim 1$ term in the denominator. The exchange constant then assumes the form,
\begin{equation}
\label{J_AB_e0short}
 J_{\rm eff}^{\! AB}({\bf R})= \frac{3J^2}{16\pi \hbar} \frac{ \! A_o^2}{v R^3}\frac{\sin^2\!{\theta_{\! AB}}}{(1+\alpha_0 \ln \frac{R}{a})^4}.
\end{equation}
The short-distance value (\ref{J_AB_e0short}) is suppressed, compared with the long-distance asymptotic (\ref{J_AB_e0}),  by the additional factor that depends weakly (logarithmically)  on $R$. To justify neglecting $\rho$ in both the numerator and the denominator, it is sufficient to have $\alpha_0\rho \ll 1$ or, equivalently,
$R \ll v/\epsilon_{o}$.

At $\rho \to 0$, the contribution of $x\sim 1$ to the integral in Eq.~(\ref{J_AB}) is the only one that matters, and leads to Eq.~(\ref{J_AB_e0short}). However, as $\rho$ increases (but still does not exceed $1$, see below), a contribution of small $x\ll 1$ might become dominant, where the term $ix$ in the numerator of the integrand can be neglected in comparison with $\alpha_0 \rho$ and where one can approximate $K_1(x)\approx 1/x$. As long as the poles of the integrand are on opposite sides of the real axis (see discussion below), the remaining integral can be calculated by the residue method,
\begin{align}
\label{J_AB_smallrho}
 J_{\rm eff}^{\! AB}({\bf R})=
&\frac{J^2}{\pi^3 \hbar}\frac{A_o^2 }{v R^3 }\Big(\frac{\epsilon_{o}R}{v \alpha_{0}}\Big)^4 \sin^2\!\theta_{AB} \times\nonumber\\
&\int\limits_{-\infty}^{\infty}  \frac{dx  }
{\left[ \left(\rho -ix\ln{\left[\frac{cR}{a|x|}\right]}
\right)^2  -\sin^2\!\theta_{AB}\right]^2}  \nonumber\\
&= \frac{\pi^2J^2}{2\pi^2 \hbar}\frac{\epsilon_o^4 v^3 R}{A_o^2\gamma^8} \frac{1}{|\sin{\theta_{\! AB}}| \ln(\frac{\alpha_{0}v}{a\epsilon_{o}})}.
\end{align}
Because $\sin{\theta_{\!AB}} \sim 1$, the typical value $x\sim 1/\ln(R/a)$ and the condition $x\ll 1$ is satisfied automatically. But to ensure that $x$ is smaller than $\alpha_0\rho$,  one must also have $\alpha_0\rho \gg 1/\ln \frac{R}{a}$ or, equivalently,
$R \gg (v/\epsilon_{o}) /\ln \frac{R}{a}$. We conclude that the exchange coupling constant is the sum of contributions (\ref{J_AB_e0short}) and (\ref{J_AB_smallrho})  in the range of distances,
$(v/\epsilon_{o}) /\ln \frac{R}{a}\ll R \ll v/\epsilon_{o}$. Within this range, with increasing $R$, the $\propto R^{-3}$ contribution gradually becomes dominated by the linear $\propto R$ term. The crossover occurs at the point where the two terms are of the same order of magnitude, at $R\sim (v/\epsilon_{o}) / (\ln \frac{R}{a})^{3/4}$.

One additional condition should be emphasized. The residue method applies only if  $\rho <\rho_0 =|\sin{\theta_{\! AB}}|$ or equivalently $R < \gamma^2 A_{0}|\sin\theta_{AB}|/\pi v \epsilon_{o}$; otherwise the poles of the integrand in Eq.~(\ref{J_AB_smallrho}) reside on the same side of the real axis. To ensure that one nonetheless  has $\alpha_0\rho \ln \frac{R}{a} \gg 1$, it is necessary that the
value $\alpha_0 \sim (\gamma/t)^2$ is not too small, i.e. that $\alpha_0 \ln \frac{R}{a} \gg 1$.

 The linear increase of the effective spin exchange coupling is caused by multiple scattering of conduction electrons off the two impurities. As the distance $\rho$ increases even further and approaches  $\rho_0=|\sin\theta_{AB}|$, a resonant enhancement of the exchange constant occurs. There, one of the energy levels of the impurities  in AB configuration crosses the Dirac point, see Eq.~(\ref{EnergyAB}). As a result, at $\rho=\rho_0$ the integrand has the singularity at $x=0$. For small values  $\rho - \rho_o$ the integral can be calculated by keeping only terms linear in $x$ in the denominator. This is justified by the fact that the integral converges at $x\sim (\rho^2 - \rho_0^2)/(\rho_0 \ln \frac{R}{a})$. In the leading logarithm approximation we obtain (see Appendix \ref{Appendix B} for details),
\begin{equation}
\label{resonantAB}
J_{\rm eff}^{AB}({\bf R})= -\frac{v J^2 \epsilon_{o}^2  }{4 \hbar \gamma^4 |R-R_0|}\frac{|\sin\theta_{AB}|}{\ln^2\!{(\frac{R_0^2}{a(R-R_0)})} }.
\end{equation}
\begin{figure}
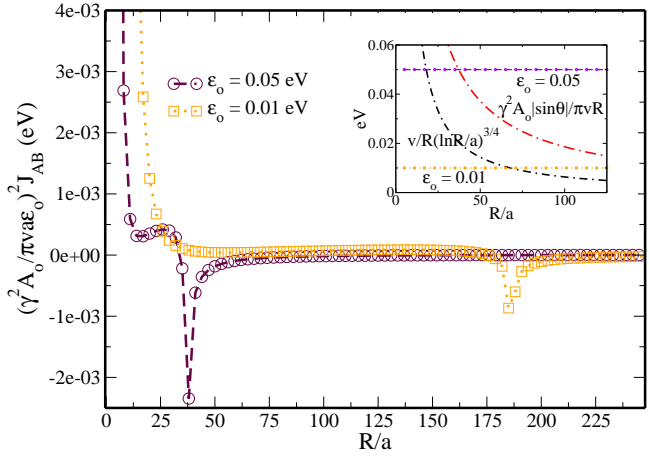

        \centering
    	\def\big{\includegraphics[scale = 0.35]{./images/reson3eV.eps}}
    	\def\little{\includegraphics[scale=0.13]{./images/insetJAB2.eps}}
    	\def\stackalignment{r}
    	\topinset{\little}{\big}{14pt}{15pt}
	\caption{Effective spin coupling $J_{AB}$ is plotted as a function of distance between the impurities $R/a$ in AB configuration for two different values of onsite energy $\epsilon_{o}$: 0.01 and 0.05 eV. The coupling constant $\gamma = 3 $eV is same for all three plots in this figure. $J_{AB}$ is scaled by dimensionless ratio $\gamma^2 A_o/\pi va \epsilon_o$. It is exact numerical plot of Eq. (\ref{J_AB}).}
	\label{fig 7}
\end{figure}

Fig.~\ref{fig 6} illustrates the dependence of the spin exchange coupling $J_{AB}$ on the distance between the impurities for different values of the onsite energy $\epsilon_{o}$ and $\gamma = 3 $ eV. The interaction changes from the weak impurity limit to strong impurity limit at short distances on decreasing value of $\epsilon_{o}$. This can be understood from the inset plot shown inside the graph, it shows the variation of $\gamma^2 A_o \ln(R/a)/\pi v R$ and $v/R$ with $R/a$.
$\epsilon_{o}$ is always greater than $\gamma^2 A_o \ln(R/a)/\pi v R$ for $\epsilon_{o} = 1.5$ eV and thus the interaction is always anti-ferromagnetic. It changes to strong impurity limit for $R \sim \gamma^2 A_o \ln(R/a)/\pi v \epsilon_{o} $ for $\epsilon_{o} = 0.5$ eV where the impurities are ferromagnetically coupled. 

Fig. \ref{fig 7} illustrates the dependence of the spin exchange coupling $J_{AB}$ on the distance between the impurities for two values of onsite energy $\epsilon_{o} = 0.01$ and $0.05$ eV and $\gamma = 3 $ eV. At small distances $R < (v/\epsilon_{o}) / (\ln \frac{R}{a})^{3/4}$, the interaction is anti-ferromagnetic and decreases with increasing $R$. On increasing distance, we see a transition from weak coupling to strong coupling. Above $R\sim (v/\epsilon_{o}) / (\ln \frac{R}{a})^{3/4}$, the interaction remains anti-ferromagnetic but increases linearly with $R$. At $R = \gamma^2 A_{0}|\sin\theta_{AB}|/\pi v \epsilon_{o}$, the spin exchange coupling becomes resonant ferromagnetic.

\subsection{AA configuration}
\begin{figure}
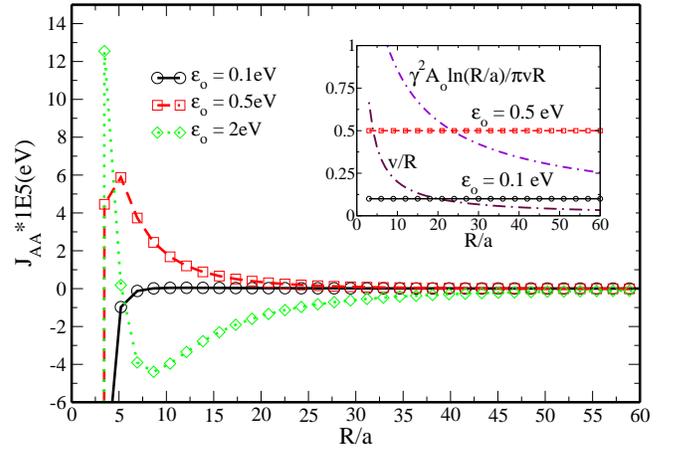

        \centering
    	\def\big{\includegraphics[scale = 0.35]{./images/JAA_w3eV.eps}}
    	\def\little{\includegraphics[scale=0.15]{./images/insetJAA.eps}}
    	\def\stackalignment{r}
    	\topinset{\little}{\big}{14pt}{16pt}
	\caption{Effective spin coupling $J_{AA}$ is plotted as a function of distance between the impurities $R/a$ in AA configuration for three different values of onsite energy $\epsilon_{o}$: 0.1, 0.5 and 2 eV. The coupling constant $\gamma = 3 $eV is same for all three plots in this figure. It is exact numerical plot of Eq. (\ref{J_AA}).}
	\label{fig 8}
\end{figure}
In the AA configuration of impurities the spin exchange coupling is given by,
\begin{align}
\label{J_AA}
&J_{\rm eff}^{AA}({\bf R})= -\frac{J^2}{\pi^3 \hbar \alpha_{0}^4}\frac{A_o^2 }{v R^3 }  \cos^2\!\theta_{AA} \times\nonumber\\
&\int\limits_{-\infty}^{\infty} dx  \frac{x^2K_0^2(|x|)(ix-\alpha_{0}\rho)^4 }
{\left[ \left(\rho + ix\Big(\frac{1}{\alpha_0}+\ln{\left[\frac{R}{a|x|}\right]\Big)}
\right)^2  + \cos^2\!\theta_{AA}x^2K_0^2(|x|)\right]^2}.
\end{align}
Similarly to AB case, we obtain the same $1/R^3$ dependence of spin exchange coupling for both small and large distances $R$. For large $\rho \gg \frac{1}{\alpha_{0}}+ \ln(\frac{R}{a})$, we can keep only the $\rho$ terms in both numerator and denominator and utilize the fact that the integral in Eq. (\ref{J_AA}) converges at $x \sim 1$. Because  $ \int_{-\infty}^{\infty} dx x^2K_0^2(|x|) =\pi^2/16$, the exchange coupling becomes,
\begin{equation}
\label{J_AA_e0}
 J_{\rm eff}^{\! AA}({\bf R}) =-\frac{J^2}{16\pi \hbar} \frac{ \! A_o^2}{v R^3}\cos^2\!{\theta_{\! AA}}.
\end{equation}
 On the other hand, for small $\rho\to 0$ and large value of $\frac{1}{\alpha_{0}}+ \ln(\frac{R}{a}) \gg 1$, we can neglect in the denominator  both  $\rho$ and the term containing $K_0(x)$:
\begin{equation}
\label{J_AA_e0short}
 J_{\rm eff}^{\! AA}({\bf R}) =-\frac{J^2}{16\pi \hbar} \frac{ \! A_o^2}{v R^3}\frac{\cos^2\!{\theta_{\! AA}}}{(1+\alpha_0 \ln \frac{R}{a})^4}.
\end{equation}
Again $x \sim 1$ values determine the integral, and thus the last expression is valid as long as $\alpha_{0} \rho \ll 1$, which is equivalent to $R \ll v/\epsilon_{o}$.

As $\rho$ increases, the contribution of small $x \ll 1$ can become important, where we can neglect $ix$ term in the numerator in comparison to $\alpha_{0} \rho$. For $\rho \ll 1$, the integral in Eq.(\ref{J_AA}) converges on $x \sim \rho$ where we can approximate the Macdonald function $K_{0}(x) \approx -\ln x$. In the leading logarithmic approximation it is typically sufficient to take the logarithms at the characteristic arguments; in this case $x\sim \rho$. But the integral thus evaluated will vanish because both poles of the integrand will lie on the same side of the $x$-axis. Thus, it is important to calculate the subleading contribution to the integral where one can no longer treat logarithms as constant. The corresponding calculation is presented in Appendix \ref{Appendix C}. In the limit of small $\epsilon_{o}$ and large distances $R$ it amounts to,
\begin{equation}
\label{J_AAsmallrho}
J_{\rm eff}^{AA}({\bf R})=  \frac{2J^2\pi v^2\epsilon_{o}^3}{3 \hbar A_0 \gamma^6} \ln\Big(\frac{v \alpha_{0}}{R \epsilon_{o}}\Big)\frac{\cos^2\theta_{AA} }{\ln^3 (\frac{v \alpha_{0}}{a \epsilon_{o}})}.
\end{equation}
We note that since $x\sim \rho$, the condition $\alpha_{0} \rho \gg x$ requires that the value of $\alpha_{0}$ is large, $\alpha_{0} \gg 1$. If this is the case,  the exchange coupling constant for $R \ll v/\epsilon_{o}$ is the sum of two contributions, Eq.~(\ref{J_AA_e0short}) and Eq.~(\ref{J_AAsmallrho}). With increasing $R$, the logarithmic contribution (\ref{J_AAsmallrho}) exceeds the  $R^{-3}$ part, Eq.~(\ref{J_AA_e0short}). This crossover occurs when the two terms becomes of the same order, at $R \sim \frac{v}{\epsilon_{o}}\ln \left(\frac{v\alpha_{o}}{a\epsilon_{o}}\right) \Big[ \frac{t}{\gamma  (\ln R/a)^2 } \Big]^{2/3}$. 


Fig. \ref{fig 8} illustrates dependence of the spin exchange coupling $J_{AA}$ on the distance between impurities for $\gamma = 3 $ eV and different values of the onsite energy $\epsilon_{o}$. The impurities are ferromagnetically coupled for large distances $R$. The change from weak to strong impurity occurs at distances, $R \sim \gamma^2 A_{o}\ln(R/a)/ \pi v \epsilon_{o}$, as seen from the inset plot. The impurities are antiferromagnetically coupled in the strong impurity limit until $\epsilon_{o} \ll v/R$ where it switches back to ferromagnetic coupling.
\section{Summary}

Two singly-occupied Anderson impurities with the energy level below the Fermi energy interact resonantly  with each other.
The interaction is facilitated by the
exchange of virtual electron-hole excitations. The sign and nature of the interaction depend on whether the impurities reside on the same sublattice or then opposite sublattices. 

For opposite sublattices both the potential part of the interaction
and the effective spin exchange coupling have resonant character when one of the energy levels of the two-impurity system passes through the Dirac points.
The potential interaction is repulsive and decays with the third power of the distance $R$ in the weak coupling limit. The resonant potential interaction  decays as the first power of the distance and is attractive. The spin-exchange part of the interaction is anti-ferromagnetic both at small and large distances. At the distances where level crosses the Dirac points, the coupling is ferromagnetic and resonantly-enhanced.

For the same sublattice, the potential part of the interaction is attractive in the weak coupling limit and repulsive in the strong coupling limit. The spin-exchange coupling  is ferromagnetic at large and small distances but reverses sign and becomes anti-ferromagnetic for intermediate distances.

\section{Acknowledgements}
We thank Oleg Starykh for helpful discussions. The work was supported by the Department of
Energy, Office of Basic Energy Sciences, Grant No. DE-
FG02-06ER46313.

\appendix
\section{Calculation of integrals involving special functions}
\label{Appendix A}
(i) The integral in Eq. (\ref{integralAB}) is of the form:
\begin{equation}
I(\beta)= -\int\limits_{-\infty}^\infty dx \frac{x^2 K^2_1\left(|x|\right)}{(x + i\beta)^2} =
-2\int\limits_{0}^\infty dx \frac{x^2 K^2_1\left(x\right)(x^2 - \beta^2)}{(x^2 + \beta^2)^2},
\end{equation}
Integrating by parts and separating the leading contribution to the integral then gives,
\begin{equation}
I(\beta) = \frac{\pi^2}{2} + 4\beta^2 \int\limits_{0}^\infty dx \frac{K_1\left(x\right)}{(x^2 + \beta^2)}\frac{d}{dx}(xK_1\left(x\right)).
\end{equation}
The main contribution to the remaining integral comes from $x \ll 1$ where we can expand $K_1\left(x\right)$ upto second order $xK_1\left(x\right) \approx 1+ x^2\ln(0.54x)/2$ and differentiate it to rewrite the integral as,
\begin{align}
I(\beta) =& \frac{\pi^2}{2} + 4\beta^2 \int\limits_{0}^\infty dx \frac{\ln(0.89x)}{(x^2 + \beta^2)}\nonumber\\
& = \frac{\pi^2}{2} + 2\pi \beta \ln{(0.89\beta)}.
\end{align}

(ii) The integral in Eq.~(\ref{integralAA}) can also be calculated in a similar way. On integration by parts we get,
\begin{equation}
I(\beta)= \int\limits_{-\infty}^\infty dx \frac{x^2 K^2_0\left(|x|\right)}{(x + i\beta)^2}
\approx 4 \int\limits_{0}^\infty dx \frac{x^2 K_0\left(x\right)}{(x^2 + \beta^2)}\frac{d}{dx}(xK_0\left(x\right)).
\end{equation}
Using $d(xK_0\left(x\right))/dx = K_{0}(x) - xK_{1}(x) $ and separating leading contribution to the integral gives,
\begin{equation}
I(\beta) = \frac{\pi^2}{2} - 4\beta^2 \int\limits_{0}^\infty dx \frac{K^2_0\left(x\right)}{(x^2 + \beta^2)} + 4\beta^2 \int\limits_{0}^\infty dx \frac{x K_0\left(x\right)K_1\left(x\right)}{(x^2 + \beta^2)}.
\end{equation}
The remaining integrals are easy to calculate by noting that only $x \sim \beta$ are important to the integral and hence for small $\epsilon$ we can approximate $K_{0}(x) \approx -\ln x$ and $K_{1}(x) \approx 1/x$ to get,
\begin{equation}
I(\beta) = \frac{\pi^2}{2} - \frac{\pi^3\beta}{2} - 2\pi \beta \ln^2\beta - 2\pi \beta \ln\beta.
\end{equation}
\section{Calculation of resonant integral}
\label{Appendix B}
To calculate the integral in Eq. (\ref{J_AB}) near resonance i.e. for small values of $\rho - \rho_{o}$ keep terms of lowest order in $x$. This is justified by the fact that most important contribution to the integral comes from small arguments $x \ll 1$. Denoting now $\xi = \rho^{2}-\rho_0^{2}$, in the leading logarithmic approximation,
\begin{align}
\label{JABappendix}
&\int\limits_{-\infty}^{\infty}  \frac{dx (\frac{ixv}{R} -\epsilon_{o})^4}{\left(\xi -2i\rho_0x (\ln{(\frac{cR}{a|x|})} \right)^2} \nonumber\\
&=-\epsilon_{o}^4 \frac{\partial}{\partial \xi} \int\limits_{0}^{\infty}  \frac{2\xi \, dx}{\xi^2 +4\rho_0^2x^2 \ln^2\!{(\frac{cR}{a|x|}
)}} =-\epsilon_{o}^4\frac{\pi}{2\rho_0|\xi|\ln^2\!{(\frac{cR}{a|\xi|}.
)}},
\end{align}
where $c$ is introduced as $\ln c  = \frac{1}{\alpha_{0}}$. In calculating the above integral we have made use of the approximation that $\frac{\xi v}{\rho_{o}\ln(\frac{cR}{a|\xi|})R}  \ll \epsilon_{o}$ thus neglecting $xv/R$ term in the first line in the numerator of Eq. (\ref{JABappendix}).
\section{Calculation of logarithmic integrals}
\label{Appendix C}
To calculate the integral in Eq. (\ref{J_AA}) for $\rho \ll 1$, we rewrite the spin exchange coupling in the following form:
\begin{multline}
\label{thirdJ_AA_StrongU}
J_{\rm eff}^{AA}({\bf R})= -\frac{J^2}{\pi^3 \hbar}\frac{A_o^2 }{v R^3 } \Big(\frac{\epsilon_{o}R}{\alpha_{0}v}\Big)^3
\int\limits_{-\infty}^{\infty}dz~c^{2}z^{2} \ln^2({B}/{|z|}) \\ \times \frac{1}{\Big[\Big\{1+i z\ln{\left(\frac{c_{1}A}{|z|}\right)}\Big\}^2+c^{2}z^{2} \ln^2(\frac{B}{|z|})\Big]^2},
\end{multline}
where we have rescaled the integration variable, $x = \rho z$ and introduced the shorthands $A =R/(a\rho)$,  $B =1/\rho$,  $c = |\cos\theta_{AA}|$ and $\ln c_{1} = \frac{1}{\alpha_{0}}$.
The above integral converges at $z \sim 1$, where we can expand the integrand up to first order in $\ln z$ to get,
\begin{align}
\label{firstI(A,B)}
I = \frac{c}{2}\frac{\partial}{\partial c} \int\limits_{-\infty}^{\infty}  \frac{dz}{\Big[\Big\{1+i z\ln{\left(\frac{c_{1}A}{|z|}\right)}\Big\}^2+c^{2}z^{2} \ln^2(\frac{B}{|z|})\Big]}\nonumber\\
=c \frac{\partial}{\partial c} \int\limits_{-\infty}^{\infty} dz \ln |z|\frac{iz(1+iz \ln c_{1}A) + c^{2}z^{2} \ln B}{[(1+iz\ln c_{1}A)^2 + c^{2}z^{2} \ln^2 B]^2}.
\end{align}
Above integral is of the form $\int_{-\infty}^\infty dz \ln|z| K(z)$, where $K(z)$ is a rational function with all its singularities located in the upper half-plane of complex $z$. Defining a new function $Q(z)$ according to $Q(z) =\int_{-\infty}^z dzK(z)$, one can use the integration by parts to obtain,
\begin{align}
\label{integral transformation}
\int\limits_{-\infty}^\infty dz \ln|z| \frac{dQ(z)}{dz} &=  -P\int\limits_{-\infty}^\infty dz \frac{Q(z)}{z} \nonumber\\ &=i\pi Q(0) =i\pi \int\limits_{-\infty}^0 dz  K(z).
\end{align}
In performing this transformation we have used that $Q(\infty)=\int_{-\infty}^\infty dzK(z)=0$ since the function $K(z)$ does not have any singularities in the lower half-plane of $z$. Additionally, to express the principal value integral in Eq.~(\ref{integral transformation}) via $Q(0)$, we observe that
\begin{equation}\label{Sokhotsky}
  \int\limits_{-\infty}^\infty dz \frac{Q(z)}{z-i0}= P\int\limits_{-\infty}^\infty dz \frac{Q(z)}{z}+i\pi Q(0) =0,
\end{equation}
as the integral in the left-hand side of Eq.~(\ref{Sokhotsky}) is zero for the already familiar reason: all its poles reside in the upper half-plane.  From Eq.~(\ref{integral transformation}) we obtain that the exchange coupling constant  (\ref{firstI(A,B)}) is expressed in terms of the following integral of a rational function,
\begin{equation}
\label{firstI(A,B)1}
I=  i\pi c\frac{\partial}{\partial c} \int\limits_{-\infty}^{0} dz \frac{iz(1+iz \ln c_{1}A) + c^{2}z^{2} \ln B}{[(1+iz\ln c_{1}A)^2 + c^{2}z^{2} \ln^2 B]^2}.
\end{equation}
The above integral can be easily calculated to get,
\begin{multline}
\label{AA strong limit}
I=  \frac{\pi}{4 c \ln^2 B} \Big[ \frac{4c^3\ln (c_{1}A/B)\ln^3 B}{(\ln^2\! c_{1}A - c^2 \ln^2 B)^2} \\ - \frac{2c\ln c_{1}A \ln B}{\ln^2 c_{1}A - c^2 \ln^2B} + \ln\Big(\frac{\ln c_{1}A + c\ln B}{\ln c_{1}A - c\ln B}\Big)\Big].
\end{multline}

\end{document}